\documentstyle[pre,twocolumn,aps]{revtex}
\begin{document}
\draft
\title{Hamiltonian reformulation and pairing of Lyapunov
exponents for Nos\'{e}-Hoover dynamics}
\author{C. P. Dettmann and G. P. Morriss}
\address{School of Physics, University of New South Wales, Sydney 2052,
Australia}
\date{To appear in Phys. Rev. E}
\maketitle
\begin{abstract}
The Nos\'{e} Hamiltonian is adapted, leading to a derivation of the 
Nos\'{e}-Hoover equations of motion which does not involve time
transformations, and in which the degree of freedom corresponding to the
external reservoir is treated on the same footing as those of the rest
of the system.  In this form it is possible to prove the conjugate
pairing rule for Lyapunov exponents of this system.
\end{abstract}

\section{Introduction}\label{int}
The Nos\'{e} Hamiltonian~\cite{N1,N2}
\begin{equation}
H_N({\bf q},s;{\mbox{\boldmath $\pi$}},p_s;\lambda)=
\sum_{i=1}^N\frac{{\mbox{\boldmath $\pi$}}_i^2}{2m_is^2}
+\varphi({\bf q})+\frac{p_s^2}{2Q}+gkT\ln s\;,\label{Nose}
\end{equation}
is used to model a system of $N$ particles interacting with a thermal
reservoir at temperature $T$, represented by the coordinate $s$ with
its conjugate momentum $p_s$.  $g$ is a constant which depends on the
number of degrees of freedom of the system.  If we interpret the time
variable $\lambda$ as a nonphysical parameter, with physical time intervals
defined by $dt=d\lambda/s$ and physical momentum ${\bf p}_i=d{\bf
q}_i/dt= \mbox{\boldmath $\pi$}_i/s$, a uniform (microcanonical)
probability measure on the full $({\bf q},{\mbox{\boldmath $\pi$}},s,p_s)$
phase space reduces to a canonical probability measure on the system
variables $({\bf q},{\bf p})$, as long as $g=3N+1$.  Thus a molecular
dynamics simulation may be used to model canonical, as well as
microcanonical ensembles, as long as the extended system dynamics is
ergodic.  Refer to Ref.~\cite{N3} for a more complete discussion and
review.  Numerical results suggest that the ergodic assumption may be
reasonable for all but the very smallest systems~\cite{PHV}.  It is
possible make a number of modifications of the time and momentum
variables in $H_N$ to obtain other dynamical systems which generate the
canonical ensemble~\cite{JB,J}.

Hoover~\cite{H} pointed out that the equations in terms of the
physical variables $({\bf q},{\bf p}, t)$  and $\zeta=p_s/Q$ take on a
particularly simple form:
\begin{eqnarray}
\dot{\bf q}_i&=&{\bf p}_i/m_i\label{Hbeg}\\
\dot{\bf p}_i&=&-\nabla_i\varphi-\zeta{\bf p}_i\\
\dot{\zeta}&=&\frac{1}{Q}\left(\sum_{i=1}^N\frac{{\bf p}^2_i}{m_i}
-gkT\right)\\
\dot{s}&=&\zeta s\;\;.\label{Hend}
\end{eqnarray}
Note that the first three equations form a closed set;
$s$ is now redundant.  In this form, it is apparent that $\zeta$ acts as
a kind of thermostat acting on the kinetic energy of the $N$-particle
system.  $\dot{\zeta}$ is proportional to the difference between the
kinetic energy and $gkT/2$, so that when the kinetic energy rises above
this value, $\zeta$ increases, raising the damping term in the
equation for $\dot{\bf p}$, and thus reducing the kinetic energy.
In the Hoover representation the equilibrium distribution is changed
because of the time scaling, however the canonical distribution in the
system variables is recovered by setting $g=3N$.

The real utility of this approach, however, is to nonequilibrium
systems.  In a nonequilibrium simulation in which energy is being pumped
through the system such a thermostat permits the system to approach a
steady state while remaining homogeneous; there are alternative approaches
involving boundaries which do not share this property~\cite{GN,GD,CL}.
Another important thermostatting method based on Gauss' principle of
least constraint~\cite{P} uses $\zeta$ as an explicit function of the
coordinates rather than a variable in its own right~\cite{EM}.  The
Nos\'{e}-Hoover and Gaussian thermostats give the same averages and time
correlation functions in the thermodynamic limit~\cite{ES}.  The
important point to note here is that any representation of a
nonequilibrium steady state must contain some reference to an external
heat reservoir.  A possible advantage of the Nos\'{e}-Hoover scheme over
the Gaussian approach is that this reservoir is explictly included as a
separate degree of freedom, so that it may be treated on a similar
footing as the rest of the system.  Recently a number of modifications and
extensions to the Nos\'{e}-Hoover method have been
proposed~\cite{MKT,R,HVR,HH}.

Lyapunov exponents (defined in Sect.~\ref{cp}) are important in the study of
nonequilibrium systems, giving information on the chaotic instability,
and providing an important link between the microscopic and macroscopic
properties, since the sum of the exponents gives the average rate of phase
space expansion, which can then be related to entropy production, and
hence transport coefficients.  The conjugate pairing rule for Lyapunov
exponents is the property that there is a constant $C$ such that for
every exponent $\lambda$, $C-\lambda$ is also an exponent, with the
possible exception of one or two exponents which are fixed to be zero by
symmetry considerations.  It was first discussed in Refs.~\cite{EM,ECM}.
The conjugate pairing rule permits the sum of
the exponents to be calculated from the largest and smallest, which are
the easiest to evaluate numerically.  Hamiltonian systems obey conjugate
pairing with $C=0$, that is, the exponents come in $\pm$
pairs~\cite{AM}.  Systems with a constant damping factor pair with $C$
proportional to this factor~\cite{D}.  Recently the conjugate pairing
rule has been shown to hold for systems containg a Gaussian
thermostat~\cite{DM}, where $C$ is minus the average value of the
thermostatting multiplier $\alpha$ (analogous to $\zeta$ here).
In this case there are two zero exponents which do not pair, which arise
from the time translation symmetry, and the conserved kinetic energy.

Note that the Lyapunov exponents depend on the variables used to define
the phase space if the equations relating different coordinate systems
involve exponential functions of time.  This means that, although it is
trivial to prove that the Lyapunov exponents obtained using the original
Nos\'{e} variables pair to zero, because the equations of motion are
derived from a Hamiltonian, it is much harder to make statements about the
Hoover variables, particularly since a different time variable is used.
Here the total phase space contraction, which is given by minus the sum of
the Lyapunov exponents, is proportional to the average of $\zeta$, which
is nonzero for a nonequilibrium steady state.  Since the sum of the
exponents is less than zero, it is clear that the exponents are quite
different to the Nos\'{e} values, for which the sum is trivially
zero.

Sect.~\ref{ham} of this paper shows how to write the Nos\'{e}
Hamiltonian in a form which treats the system and reservoir variables on
equal footing, leading to a proof of the conjugate pairing rule in
Sect.~\ref{cp} , along the lines of the proof in Ref.~\cite{DM}.

\section{Unified Hamiltonian formalism}\label{ham}
In the form given by Nos\'{e} (Eq.~\ref{Nose}), the reservoir variable
$s$ is treated quite differently to the other coordinates in that the
system kinetic term is divided by $s^2$ wheras the reservoir kinetic
term is not.  In addition, the time $\lambda$ corresponding to the
Hamiltonian does not correspond to physical time.  In this section $H_N$
is transformed to alleviate both of these deficiencies.

The unified form of the Hamiltonian is obtained by transforming to a new
coordinate $\sigma=\ln s$.  This type of transformation is described in
Sect.~9.2 of Ref.~\cite{G}, and uses a generating function of the form
\begin{equation}
F_2({\bf q},s;\mbox{\boldmath $\pi$},p_\sigma;\lambda)=
\sum_{i=1}^N{\bf q}_i\cdot\mbox{\boldmath $\pi$}_i+p_\sigma\ln s\;\;,
\end{equation}
leading to a new momentum $p_\sigma=sp_s$, and transformed Hamiltonian
\begin{eqnarray}
H_T({\bf q},\sigma;\mbox{\boldmath $\pi$},p_\sigma;\lambda)&=&
\frac{e^{-2\sigma}}{2}\left(\sum_{i=1}^N
\frac{\mbox{\boldmath $\pi$}_i^2}{m_i}+\frac{p_\sigma^2}{Q}\right)
\nonumber\\&+&\varphi({\bf q})+gkT\sigma.
\end{eqnarray}
Note that the form of the potential is also simpler in this
representation.  The masses may also be scaled out: Construct $3N+1$
dimensional vectors ${\bf X}=({\bf q}_i\sqrt{m_i},\sigma\sqrt{Q})$ and
${\bf P}=(\mbox{\boldmath $\pi$}_i/\sqrt{m_i},p_\sigma/\sqrt{Q})$, and write
$\Phi({\bf X})=-\sigma$ and $\phi({\bf X})=\varphi({\bf q})+gkT\sigma$.
The Hamiltonian may then be written in a unified form as
\begin{equation}
H_U({\bf X};{\bf P};\lambda)=\frac{{\bf P}^2}{2}e^{2\Phi({\bf X})}
+\phi({\bf X})\;\;.
\end{equation}

The final transformation which eliminates the need for an unphysical
time variable follows similarly to the Hamiltonian for the Gaussian
thermostat~\cite{DM2}.  First add a constant to $\phi$ so that the
initial (and hence at all times) value of $H_U$ is zero.  It is easily
verified that multiplying a zero Hamiltonian by an arbitrary function
scales the time, but has no other effect on the equations of motion.
Thus the final form of the Hamiltonian is
\begin{equation}
H_F({\bf X};{\bf P};t)=\frac{{\bf P}^2}{2}e^{\Phi({\bf X})}
+\phi({\bf X})e^{-\Phi({\bf X})}\;\;.
\end{equation}
This transformation is equivalent to multiplying $H_N$ in
Eq.~(\ref{Nose}) by $s$, which is a particularly simple method of
generating Eqs.~(\ref{Hbeg}-\ref{Hend}) without the use of unphysical
time variables.

It is also convenient to introduce a few more $3N+1$ dimensional
vectors,
\begin{eqnarray}
{\bf V}&=&{\bf P}e^{\Phi}\\
{\bf F}&=&-\nabla\Phi\\
{\bf f}&=&-\nabla\phi\;\;,
\end{eqnarray}
in terms of which the condition $H_F=0$ becomes
\begin{equation}
\frac{{\bf V}^2}{2}+\phi=0\;\;,\label{cons}
\end{equation}
and Hamilton's equations of motion reduce to
\begin{eqnarray}
\dot{\bf X}&=&{\bf V}\label{eombeg}\\\label{eomend}
\dot{\bf V}&=&{\bf V}^2{\bf F}-{\bf F}\cdot{\bf V}{\bf V}+{\bf f}\;\;.
\end{eqnarray}

As an example, let us consider the Nos\'{e}-Hoover oscillator of
Ref.~\cite{PHV}.  For a single particle in a one dimensional harmonic
oscillator potential, $\varphi=m\omega^2q^2/2$.  Then the Hamiltonian becomes
\begin{eqnarray}
H_F(q,\sigma;\pi,p_\sigma;t)&=&\left(\pi^2/(2m)+p_\sigma^2/(2Q)\right)
e^{-\sigma}\nonumber\\
&+&\left(m\omega^2q^2/2+gkT\sigma\right)e^\sigma\;\;,
\end{eqnarray}
and the vectors we introduced above are
\begin{eqnarray}
{\bf X}&=&(q\sqrt{m},\sqrt{Q}\ln s)\\
{\bf V}&=&(\pi/(s\sqrt{m}),p_\sigma/(s\sqrt{Q}))
=(p/\sqrt{m},\sqrt{Q}\zeta)\\
{\bf F}&=&(0,1/\sqrt{Q})\\
{\bf f}&=&(-\omega^2q\sqrt{m},-gkT/\sqrt{Q})\;\;.
\end{eqnarray}
In this form it is straightforward to show that
Eqs. (\ref{eombeg},\ref{eomend}) are equivalent to Hoover's form of the
equations, Eqs.~(\ref{Hbeg}-\ref{Hend}).  The decoupling of the $s$
(or $\sigma$) equation occurs because $\bf F$ and $\bf f$ are independent
of $s$.

\section{Conjugate pairing}\label{cp}
The equations of motion given in the previous section are
now in a form suitable for a proof of the conjugate pairing rule for
Lyapunov exponents.  Note that the coordinates used are the same (apart
from constants related to the masses) as the Nos\'{e}-Hoover thermostat,
except that $s$ is replaced
by $\sigma$.  The components of $V$ are proportional to the physical
momentum $\bf p$ and $\zeta$.  Since the equations do not depend on
$\sigma$, the Lyapunov exponents obtained contain one extra zero
exponent, but otherwise are the same as the other $6N+1$ equations
considered separately.

The argument given here closely follows Ref.~\cite{DM}, so the
presentation will be correspondingly brief.  It is convenient to
group $\bf X$ and $\bf V$ together to form a point $\Gamma$ 
in $6N+2$ dimensional phase space.  Time dependent matrices $T$ and $L$
are defined, giving the infinitesimal and finite evolution of linear
perturbations $\delta\Gamma$ as follows.
\begin{eqnarray}
\delta\dot{\Gamma}(t)&=&T(t)\delta\Gamma(t)\\
\delta\Gamma(t)&=&L(t)\delta\Gamma(0)
\end{eqnarray}
The Lyapunov exponents are defined as the logarithms of the eigenvalues
of $\Lambda$, where
\begin{equation}
\Lambda=\lim_{t\rightarrow\infty}\left(L^T(t)L(t)\right)^{1/(2t)}
\end{equation}

Two of the Lyapunov exponents are zero due to the fact that
perturbations along the flow simply add a constant to the time, and the
conservation of the value of the Hamiltonian, Eq.~(\ref{cons}),
which is the total (scaled) energy of the system and reservoir.  Of
course the energy of the system alone is not conserved, as it maps out a
canonical distribution.

The conjugate pairing clearly does not include these two exponents, so
perturbations along the flow, and which alter the value of the
Hamiltonian must be eliminated before the pairing is apparent.  This is
achieved by considering $6N$ perturbations, none of which are along the
flow or alter the total energy.  They are measured with respect to a
basis in a $6N$ dimensional subspace which rotates with the trajectory in
order to enforce these properties.  In particular $\delta{\bf X}$ is
always perpendicular to $\bf V$ (in $3N+1$ dimensional coordinates),
and $\delta{\bf V}$ has a component parallel to $\bf V$ which is fixed
by energy conservation.  As the perturbed trajectory evolves,
it will always have the same conserved energy, but $\delta{\bf X}$ may
not remain perpendicular to $\bf V$.  The perturbed trajectory will,
however cross the $6N$ dimensional space at some time $t^{'}$ different
to $t$, so the above conditions may be enforced by allowing the
perturbed trajectory to evolve at a rate different to the original.
The Lyapunov exponents obtained using this approach are the same as for
the full $6N+2$ dimensional space, with the exception of the two zeros.
These issues are discussed in more detail in Ref.~\cite{DM}, the only
difference being that here the phase space is not compact.  The
arguments follow through exactly the same, however, if ergodicity is
assumed, as it is when deriving the canonical distribution
(Sect.~\ref{int}). 

The first step is to choose an orthonormal basis in $3N+1$ dimensional space, 
$\{{\bf e}_0,{\bf e}_i\}$, with $i$ ranging from $1$ to $3N$, as it will
henceforth.  ${\bf e}_0$ is parallel to ${\bf V}$,
\begin{equation}
{\bf V}=V{\bf e}_0\;\;,
\end{equation}
with $V=|{\bf V}|$ and all the ${\bf e}_i$ are perpendicular to $\bf V$.
This basis is used for both $\bf X$ and $\bf V$ subspaces.  The equations of
motion~(\ref{eombeg}-\ref{eomend}) determine the time evolution of $V$
and ${\bf e}_0$,
\begin{eqnarray}
\dot{V}&=&{\bf f}\cdot{\bf e}_0\label{Veom}\\\label{e0eom}
\dot{\bf e}_0&=&\sum_i(V{\bf F}+V^{-1}{\bf f})\cdot{\bf e}_i{\bf e}_i
\;\;.\end{eqnarray}
The equations of motion for the other basis vectors are somewhat
arbitrary, but the natural choice which preserves the orthonormal
character is parallel transport along the trajectory,
\begin{equation}\label{eieom}
\dot{\bf e}_i=-(V{\bf F}+V^{-1}{\bf f})\cdot{\bf e}_i{\bf e}_0\;\;.
\end{equation}

The perturbed trajectory subject to the above conditions then becomes
\begin{eqnarray}
{\bf X}^{'}&=&{\bf X}+\sum_i\delta X_i{\bf e}_i\label{X'def}\\
{\bf V}^{'}&=&{\bf V}+V^{-1}\sum_i{\bf f}\cdot{\bf e}_i\delta X_i
+\sum_i\delta V_i{\bf e}_i\;\;,\label{V'def}
\end{eqnarray}
with equations of motion
\begin{eqnarray}
\frac{d}{dt^{'}}{\bf X}^{'}&=&{\bf V}^{'}\label{X'eom}\\
\frac{d}{dt^{'}}{\bf V}^{'}&=&{\bf V}^{'2}{\bf F}^{'}-
{\bf F}^{'}\cdot{\bf V}^{'}{\bf V}^{'}+{\bf f}^{'}\;\;.\label{V'eom}
\end{eqnarray}
Here, ${\bf F}^{'}$ and ${\bf f}^{'}$ are the values at the perturbed
positions, that is,
\begin{eqnarray}
{\bf F}^{'}&=&{\bf F}+\sum_i\delta X_i\nabla_i{\bf F}\label{F'def}\\
{\bf f}^{'}&=&{\bf f}+\sum_i\delta X_i\nabla_i{\bf f}\;\;.\label{f'def}
\end{eqnarray}

Substituting Eqs.~(\ref{X'def},\ref{V'def},\ref{F'def},\ref{f'def})
into Eqs.~(\ref{X'eom},\ref{V'eom}), ignoring quadratic perturbations,
simplifying with the help of
Eqs.~(\ref{eombeg},\ref{eomend},\ref{Veom}-\ref{eieom}), and
taking components in the directions of ${\bf e}_0$ and the ${\bf e}_i$
leads to $6N+2$ equations.  One of these is not independent of the
others, due to energy conservation.  One relates $t^{'}$ and $t$, and
the remaining $6N$ determine the evolution of the perturbations:
\begin{eqnarray}
\frac{dt}{dt^{'}}&=&1+\sum_i({\bf F}+2V^{-2}{\bf f})\cdot{\bf e}_i
\delta X_i\\
\delta\dot{X}_i&=&\delta V_i\\
\delta\dot{V}_i&=&\sum_j\delta X_j{\bf e}_j\cdot
(V^2\nabla{\bf F}+\nabla{\bf f}-V^2{\bf F}{\bf F}\nonumber\\
&-&3V^{-2}{\bf f}{\bf f} -{\bf F}{\bf f}-{\bf f}{\bf F})\cdot{\bf e}_i
-V{\bf F}\cdot{\bf e}_0\delta V_i\;\;.
\end{eqnarray}
Note that $V{\bf F}\cdot{\bf e}_0$ is simply $\zeta$, and the terms
containing gradients of forces are symmetric, since $\bf F$ and $\bf f$
are derived from potentials.  From these
equations, the infinitesimal evolution matrix $T$ for the restricted
$6N$ dimensional space may be read off as
\begin{equation}
T=\left(\begin{array}{cc}0&I\\M&-\zeta I\end{array}\right)\;\;,
\end{equation}
where each of the elements are $3N\times3N$ submatrices.  $M$ is
symmetric because $\bf F$ and $\bf f$ have been derived from a
potential, and $0$ and $I$ are the zero and unit matrices, respectively.
$T$ satisfies the equation
\begin{equation}
T^TJ+JT=-\zeta J\;\;,\label{inf}
\end{equation}
where $J$ is given by
\begin{equation}
J=\left(\begin{array}{cc}0&I\\-I&0\end{array}\right)\;\;.
\end{equation}

From this point the analysis is exactly the same as Ref.~\cite{DM}.
Eq.~(\ref{inf}) leads to similar relations for $L$ and $L^TL$, and hence
the eigenvalues of $\Lambda$.  The end result is that for each exponent
$\lambda$, $C-\lambda$ is also an exponent, with
\begin{equation}
C=-<\!\zeta\!>_t\;\;,
\end{equation}
that is, minus the time average of $\zeta$.  Thus the conjugate pairing
rule holds in the $({\bf q},\sigma,{\bf p},\zeta)$ variables, with the
exception of the two zero exponents.  As noted before, none of the
equations of motion depend on $\sigma$, so that, omitting the $\sigma$
equation gives the same exponents (which pair, as shown above), with
only one zero exponent.

Conjugate pairing has now been shown for Gaussian and Nos\'{e}-Hoover
thermostats which act on the kinetic energy.
Numerical simulations~\cite{I} suggest that the conjugate pairing rule
holds also for Gaussian thermostats which keep the internal energy
(rather than the kinetic energy) constant.

\section*{Acknowledgements}
We would like to thanks W. G. Hoover and F. Zhang for
helpful discussions.  This work was supported by the Australian Research
Council.

\end{document}